\documentclass[letterpaper, 10 pt, conference]{ieeeconf}  

\IEEEoverridecommandlockouts  

\usepackage{fix-cm}
\usepackage{etex}

\usepackage{dblfloatfix}

\usepackage{nag}

\makeatletter
\@ifpackageloaded{xcolor}{}{%
\usepackage[table,x11names,dvipsnames,svgnames]{xcolor}%
}
\makeatother

\usepackage{colortbl}

\usepackage{graphicx}
\usepackage{wrapfig}

\definecolorset{RGB}{lyft}{}{Red,194,39,36;Sunset,202,53,33;Orange,205,68,20;Amber,200,117,42;Yellow,242,169,52;Citron,186,188,44;Lime,112,159,33;Green,56,139,31;Mint,45,118,56;Teal,52,133,135;Cyan,60,132,202;Blue,55,94,248;Indigo,64,13,247;Purple,115,42,248;Pink,176,25,145;Rose,176,32,75}

\usepackage{cite}

\usepackage{microtype}

\usepackage{array}
\usepackage{multirow}
\usepackage{booktabs}
\usepackage{makecell} 

\ifcsname labelindent\endcsname
\let\labelindent\relax
\fi
\usepackage[inline]{enumitem}

\usepackage{subfig}

\newenvironment{lenumerate}[2][]
{\begin{enumerate}[label=(#2\arabic*),leftmargin=0.2in,itemindent=0.15in,#1]}
{\end{enumerate}}

\setlist*[enumerate,1]{label={\itshape\arabic*)}}

\makeatletter
\newcommand{\paragraphswithstop}{%
\let\copyparagraph\paragraph%
\renewcommand\paragraph[1]{\copyparagraph{##1.}}%
}
\makeatother

\usepackage[framemethod=tikz]{mdframed}

\usepackage{suffix}

\usepackage{environ}

\makeatletter
\newsavebox{\boxifnotempty}
\newcommand{\displayifnotempty}[3]{\sbox\boxifnotempty{#2}\setbox0=\hbox{\usebox{\boxifnotempty}\unskip}%
\ifdim\wd0=0pt
\else
 #1\usebox{\boxifnotempty}#3%
\fi%
}
\newcommand{\ifempty}[2]{\setbox0=\hbox{#1\unskip}%
\ifdim\wd0=0pt%
 #2%
\fi%
}
\newcommand{\ifnotempty}[2]{\setbox0=\hbox{#1\unskip}%
\ifdim\wd0>0pt%
 #2%
\fi%
}
\makeatother

\usepackage{algpseudocode}
\usepackage{algorithm}

\usepackage{scrlfile}

\makeatletter
\newcommand*\newstoreddef[1]{
  \BeforeClosingMainAux{%
    \immediate\write\@auxout{%
      \string\restoredef{#1}{\csname #1\endcsname}%
    }%
  }%
}
\newcommand*{\restoredef}[2]{
  \expandafter\gdef\csname stored@#1\endcsname{#2}%
}
\newcommand*{\storeddef}[1]{
  \@ifundefined{stored@#1}{0}{\csname stored@#1\endcsname}%
}
\makeatother

\usepackage{pageslts}
\NewEnviron{tee}{\BODY\typeout{Marker Tee [start] ^^J \BODY ^^JMaker Tee [end]}}

\input{preamble/math}

\newcommand{\bmat}[1]{\begin{bmatrix}#1\end{bmatrix}}

\newcommand{\transpose}{^\mathrm{T}}

\newcommand{\subjectto}{\textrm{subject to }}

\providecommand{\cC}{\mathcal{C}}

\providecommand{\cE}{\mathcal{E}}
\providecommand{\cF}{\mathcal{F}}
\providecommand{\cG}{\mathcal{G}}

\providecommand{\cL}{\mathcal{L}}

\providecommand{\cU}{\mathcal{U}}
\providecommand{\cV}{\mathcal{V}}

\providecommand{\cX}{\mathcal{X}}

\usepackage{units}

\newcommand{\newcolorlabel}[2]{%
  \expandafter\newcommand\csname #1\endcsname[1]{%
    \colorbox{#2}{\color{white}\textsf{\textbf{##1}}}}%
}

\newcommand{\newcommenter}[2]{%
  \expandafter\newcommand\csname #1\endcsname[1]{%
    \fcolorbox{#2}{#2}{\color{white}\textsf{\textbf{#1}}}
    {\color{#2}##1}}%
  \expandafter\newcommand\csname at#1\endcsname{%
    \fcolorbox{#2}{#2}{\color{white}\textsf{\textbf{@#1}}}
    {\color{#2}}}%
  \expandafter\newcommand\csname #1hl\endcsname[2]{%
    \colorbox{#2}{\color{white}\textsf{\textbf{#1}}}\sethlcolor{Azure2}\hl{##2}~%
    \expandafter\ifx\csname commentarrow\endcsname\relax$\leftarrow$\else \commentarrow[#2]\fi~%
    {\color{#2}##1}}%
  \expandafter\newcommand\csname #1st\endcsname[2]{%
    \colorbox{#2}{\color{white}\textsf{\textbf{#1}}}\sout{##2}~%
    \expandafter\ifx\csname commentarrow\endcsname\relax$\leftarrow$\else \commentarrow[#2]\fi~%
    {\color{#2}##1}}%
}
\newcommenter{TODO}{DodgerBlue1}
\newcommenter{rtron}{Green3}

\usepackage{comment}

\usepackage{pdfcomment}

\usepackage{soul}

\usepackage[normalem]{ulem}

\usepackage{csquotes}

\usepackage{tikz}
\usetikzlibrary{calc}
\usetikzlibrary{matrix}
\usetikzlibrary{chains}
\usetikzlibrary{shapes.geometric}
\usetikzlibrary{arrows.meta}
\usetikzlibrary{decorations.pathreplacing}
\usetikzlibrary{backgrounds}

\tikzset{
  dim above/.style={to path={\pgfextra{
        \pgfinterruptpath
        \draw[>=latex,|->|] let
        \p1=($(\tikztostart)!1.5em!90:(\tikztotarget)$),
        \p2=($(\tikztotarget)!1.5em!-90:(\tikztostart)$)
        in(\p1) -- (\p2) node[pos=.5,sloped,above]{#1};
        \endpgfinterruptpath
      }
    }
  },
  dim double above/.style={to path={\pgfextra{
        \pgfinterruptpath
        \draw[>=latex,|->|] let
        \p1=($(\tikztostart)!3em!90:(\tikztotarget)$),
        \p2=($(\tikztotarget)!3em!-90:(\tikztostart)$)
        in(\p1) -- (\p2) node[pos=.5,sloped,above]{#1};
        \endpgfinterruptpath
      }
    }
  },
  dim below/.style={to path={\pgfextra{
        \pgfinterruptpath
        \draw[>=latex,|->|] let 
        \p1=($(\tikztostart)!-1em!-90:(\tikztotarget)$),
        \p2=($(\tikztotarget)!-1em!90:(\tikztostart)$)
        in (\p1) -- (\p2) node[pos=.5,sloped,below]{#1};
        \endpgfinterruptpath
      }
    }
  },
}

\tikzset{
    right angle quadrant/.code={
        \pgfmathsetmacro\quadranta{{1,1,-1,-1}[#1-1]}     
        \pgfmathsetmacro\quadrantb{{1,-1,-1,1}[#1-1]}},
    right angle quadrant=1, 
    right angle length/.code={\def\rightanglelength{#1}},   
    right angle length=2ex, 
    right angle symbol/.style n args={3}{
        insert path={
            let \p0 = ($(#1)!(#3)!(#2)$) in     
                let \p1 = ($(\p0)!\quadranta*\rightanglelength!(#3)$), 
                \p2 = ($(\p0)!\quadrantb*\rightanglelength!(#2)$) in 
                let \p3 = ($(\p1)+(\p2)-(\p0)$) in  
            (\p1) -- (\p3) -- (\p2)
        }
    }
}

\newcommand{\pgfextractangle}[3]{%
    \pgfmathanglebetweenpoints{\pgfpointanchor{#2}{center}}
                              {\pgfpointanchor{#3}{center}}
    \global\let#1\pgfmathresult  
}

\usetikzlibrary{shapes.arrows}
\newcommand{\commentarrow}[1][Azure4]{\tikz[baseline=-3pt]{\node[shape border uses incircle, fill=#1,rotate=180,single arrow, inner sep=1pt, minimum size=6pt, single arrow head extend=2pt]{};}}

\tikzset{ax/.style={-latex,line width=2pt}}

\tikzset{camera/.style={fill=Sienna1,fill opacity=0.5},%
image plane/.style={draw=RoyalBlue3,line width=2pt}}

\newcommenter{todo}{Firebrick1}

\usepackage{amsmath}
\usepackage{bm}
\usepackage{graphicx}
\usepackage{algpseudocode}
\usepackage{algorithm}
\DeclareGraphicsExtensions{.pdf,.png,.jpg}
\usepackage{alphabeta}
\usepackage{fixltx2e}

\usepackage{cleveref}
\title{\LARGE \bf
Control-Based Planning over Probability Mass Function Measurements via Robust Linear Programming}
\author{Mehdi Kermanshah$^{1*}$, Calin Belta$^1$ and Roberto Tron$^{1}$
  \thanks{This work was supported by ONR MURI N00014-19-1-2571 ``Neuro-Autonomy: Neuroscience-Inspired Perception, Navigation, and Spatial Awareness''}
  \thanks{$^{1}$Mehdi Kermanshah, Calin Belta and Roberto Tron are with the Department of Mechanical Engineering,
    Boston University, 730 Commonwealth Ave, MA 02215, United States
    {\tt\small \{mker,cbelta,tron\}@bu.edu}}%
}%

\begin{document}

\maketitle 
\thispagestyle{empty}
\pagestyle{empty}
\newcommenter{calin}{blue}
\newcommenter{mehdi}{red}


\begin{abstract}
We propose an approach to synthesize linear feedback controllers for linear systems in polygonal environments. Our method focuses on designing a robust controller that can account for uncertainty in measurements. Its inputs are provided by a perception module that generates probability mass functions (PMFs) for predefined landmarks in the environment, such as distinguishable geometric features. We formulate an optimization problem with Control Lyapunov Function (CLF) and Control Barrier Function (CBF) constraints to derive a stable and safe controller. Using the strong duality of linear programs (LPs) and robust optimization, we convert the optimization problem to a linear program that can be efficiently solved offline. At a high level, our approach partially combines perception, planning, and real-time control into a single design problem. An additional advantage of our method is the ability to produce controllers capable of exhibiting nonlinear behavior while relying solely on an offline LP for control synthesis.
 \end{abstract}

\section{Introduction}
 Using cameras as sensors for robotic applications has become ubiquitous in recent years, due to their relatively inexpensive cost and the amount of information they provide. Several algorithms have been developed to use cameras for perception in robotic control \cite{malis2002survey,rahmatizadeh2018vision, zhao2016review}. These algorithms can be broadly categorized into two main paradigms: \emph{model-based methods} and \emph{data-driven end-to-end control} \cite{hou2013model}. 
Model-based approaches involve processing the sensor data to determine the system state, while end-to-end methods map sensor outputs to control actions directly.

Model-based approaches traditionally divide the pipeline into four blocks: perception, filtering and localization, planning, and real-time control. Sensors such as cameras, lidar, and range sensors capture environmental data in the perception phase. This data is then processed to reduce the uncertainty of the measurement via filtering methods such as Kalman filtering \cite{kalman1960new} and its variants \cite{sarkka2013bayesian}, as well as particle filtering \cite{gordon1993novel}. Typically, the information is reduced to state or state plus covariance estimates. 
Planning and control involve steering a robot from a starting point to a destination while avoiding obstacles.  Planning algorithms find a nominal trajectory from a start to a goal using methods such as rapidly exploring random trees (RRT) \cite{lavalle2001rapidly}, probabilistic road maps (PRM) \cite{kavraki1996probabilistic}, and their optimal variants RRT* and PRM* \cite{karaman2011sampling}. 
Nominal trajectories are then tracked by low-level real-time controllers. 
Consequently, any misalignment between high-level planning 
and low-level controllers, due to imperfection mapping in the planning stage or the disturbances and noise experienced by the real-time controller, can result in a significant deviation from the nominal trajectory.

Real-time control methods provide control inputs at each step. One way to achieve this is through constrained optimization \cite{camacho2013model}. Control Lyapunov \cite{lyapunov1992general} and barrier \cite{ames2016control} functions are commonly used to ensure stability and safety in this framework. This method was shown to perform well in different applications such as adaptive cruise control \cite{ames2014control}, multi-agent systems \cite{wang2017safety}, and bipedal robots \cite{hsu2015control}. In these methods, real-time onboard computation is required.
These approaches can fail in some scenarios due to undesirable local equilibria or infeasible optimization\cite{reis2020control}. These methods have also been extended to incorporated model uncertainty \cite{cohen2023adaptive,jankovic2018robust, gurriet2019realizable}, measurement uncertainty \cite{dean2021guaranteeing, wang2022observer}, and both 
model and measurement uncertainty \cite{zhang2022control, agrawal2022safe, garg2021robust}. With specific relevance to this paper, Measurement Robust Control Barrier Functions (MR-CBF) \cite{dean2021guaranteeing} deal with bounded uncertainty in the states, and seek to satisfy conventional CBF constraints for all potential observations, particularly for vision-based control. The authors over-approximate CBF constraints to ensure safety for the worst-case scenario. Comparable bounds have been applied to Input-to-State Stable (ISS) observers \cite{agrawal2022safe}.

 Data-driven end-to-end (learning-based) methods have demonstrated promising performance \cite{nvidia,offself}, but they require substantial labeled datasets and computationally intensive training.
Reinforcement learning \cite{sutton2018reinforcement} avoids the need for labeled data but requires the design of a reward function, which can be challenging. More importantly, these methods generally suffer from a lack of interpretability and formal safety guarantees. 

    In this paper, we propose an approach that bridges these paradigms, incorporating a learning-based module for perception and a model-based control design that accounts for measurement uncertainty. We build upon our earlier work \cite{mahroo}, which introduced a real-time control approach to design safe and stable control fields for polygonal environments, while tightly coupling planning and low-level controllers 
    the method demonstrated significant robustness to map deformation. In our previous work, we considered measurements that were only deterministic \cite{mahroo, mahroorrt} or Gaussian with bounded variance \cite{chenfei}. In this paper, we consider measurement modeled by full probability mass functions (PMF).


In our method, we divide the environment into convex cells and design a controller for each cell. Each controller takes PMFs of pre-specified landmarks in the environment as input. In practice, such measurements can be obtained by deep Convolutional Neural Networks \cite{deepCNN}). Our method explicitly takes into account the (bounded) uncertainty captured by the PMF and calculates the offline controller using robust Linear Programming.

As opposed to the majority of CLF-CBF controllers, our approach targets offline synthesis instead of online optimization of a control signal; this has the benefit of requiring less computation during deployment (or allowing faster control update rates) and certifies that, under the nominal condition, the control action is always well defined (i.e., we do not have to consider infeasibility during deployment). Furthermore, our formulation effectively handles nonlinear measurements with uncertainty without resorting to the introduction of additional terms that make the CLF and CBF more conservative (and, hence, more likely to be infeasible).

With respect to our previous work on CLF-CBF control-based planning\cite{mahroo}, we consider measurement models and noise distributions that are more aligned with the output of learning-based methods, going beyond standard linear point statistics such as plus Gaussian noise assumptions\cite{wang2021chance}. As a result, our method displays nonlinear behavior despite requiring only convex linear optimization.

Our approach enables combining planning, real-time control, and (partial) perception into a single design procedure, whereas these are typically considered separate steps.
With respect to fully learning-based end-to-end methods, we offer formal guarantees on safety and stability, with interpretable robustness margins.



\section{Notation and Preliminaries}
\label{sec:preliminareis}

The $i$-th element of a vector $v$ is denoted as $[v]_i$. A unit vector $e_j$ is defined as a vector with all entries equal to zero, except for $[e]_j=1$. The vector of all ones is denoted as $\boldsymbol{1}$. We use $\mathbb{N}_{\leq n}$ to represent the set $\{x \in \mathbb{N} \mid x \leq n\}$, $A^{\vee} \in \mathbb{R}^{nm \times 1}$ for the vectorized version of matrix $A \in \mathbb{R}^{n \times m}$, and $A \odot B $ for element-wise multiplication of matrices $A$ and $B$.  The Lie derivative of function $h$ with respect to a linear vector field $Ax$ and matrix $B$ are denoted as $\cL_{Ax} h(x)$ and $\cL_{B}$ respectively. We use $\mathbb{E}_p[X]$ to denote the expected value of a  discrete random variable $X$ with distribution $P$. 

\subsection{Dynamic Model}

In this study, we consider a first-order linear time-invariant (LTI) system that operates in a $d$-dimensional workspace. The system's dynamics are described as follows:
\begin{equation}
\label{eq:dynamics}
 \dot{x} = Ax+Bu,
\end{equation}
where $x \in \cX \subset \mathbb{R}^{d}$ is the state of the system, $u\in \cU \subset \mathbb{R}^{n_u}$ is the vector of control inputs ($\cU$ is the control constraint set)  and ($A \in \mathbb{R}^{d \times d},B \in \mathbb{R}^{d \times n_u}$) is a controllable pair of matrices. 

\subsection{Control Lyapunov and Barrier Functions}

We use Control Lyapunov Functions (CLFs) and  Control Barrier Functions (CBFs) to ensure the stability and safety of our controller \cite{ames2016control}. In this section, we review these concepts for the particular case of the linear dynamics \eqref{eq:dynamics}.


\begin{definition}(Control Lyapunov Function \cite{ames2014control}).
A continuously differentiable positive-definite function  $V(x) :\mathbb{R}^n \rightarrow \mathbb{R}^+$ 
is a control Lyapunov function (CLF) for system \eqref{eq:dynamics} if it satisfies:
   \begin{equation}
   \label{eq:clfdef}
        \cL_{Ax} V(x) +\cL_{B}V(x)u + \alpha_v V(x) \leq 0
   \end{equation}
\end{definition}
Where $\alpha_v$ is a positive constant.
\begin{theorem}(\cite{ames2014control}). 
Given system \eqref{eq:dynamics}, if there exists a CLF $V(x)$ satisfying \eqref{eq:clfdef}, then any controller $u(x) \in K_{clf}$ (x) asymptotically stabilizes the system to $x =0$. Where
\begin{equation}
    K_{clf}(x) = \{ u \in \cU|  \cL_{Ax} V(x) +\cL_{B}V(x)u + \alpha_v V(x) \leq 0 \}
\end{equation}

\end{theorem}





\begin{definition} (Forward Invariance).
    Consider set $\cC \subset \mathbb{R}^d$ and initial condition $x_0$. The set $\cC$ is forward invariant for system \eqref{eq:dynamics} if $x_0 \in \cC$ then $x(t) \in \cC, \forall t \geq 0$.
\end{definition}
If a set $\cC$ can be rendered forward invariant, then system \eqref{eq:dynamics} is safe with respect to $\cC$. 

We describe set $\cC$ with a sufficiently smooth function $h(x): \mathbb{R}^d \rightarrow \mathbb{R}$ as follows:
\begin{equation}
\begin{aligned}
\label{eq:h}
     \cC  &= \{ x \in \mathbb{R}^n | h(x) \geq 0\} \\
      \partial{\cC}  &= \{ x \in \mathbb{R}^n | h(x) = 0\} \\
       Int(\cC)  &= \{ x \in \mathbb{R}^n | h(x) > 0\} \\
\end{aligned}
\end{equation} 
 \begin{definition}(Control barrier function \cite{ames2014control}). The function $h(x)$ is control barrier function (CBF) in \eqref{eq:h} for system \eqref{eq:dynamics} if there exists a constant scalar $\alpha_h \in \mathbb{R}^+$ such that
 \begin{equation}
\label{eq:cbfdef}
\mathcal{L}_{Ax} h(x) + \mathcal{L}_{B}h(x)u + \alpha_h h(x) \geq 0 \quad \forall x \in \mathcal{C}.
\end{equation}
     
 \end{definition}

\begin{theorem}(\cite{ames2014control}). Suppose $\cC$ is defined as in \eqref{eq:h}. If $h$ is a CBF on $\cC$ and $\frac{\partial{h}}{\partial{x}}(x) \neq 0$, $\forall x \in \partial{\cC}$ then any controller $u(x) \in K_{cbf}(x)$ for system \eqref{eq:dynamics} renders $\cC$ forward invariant, where
\begin{equation}
    K_{cbf}(x) = \{u \in \cU | \mathcal{L}_{Ax} h(x) + \mathcal{L}_{B}h(x)u + \alpha_h h(x) \geq 0\}
\end{equation}
\end{theorem}
\begin{figure}
    \includegraphics[height=0.29\textwidth]{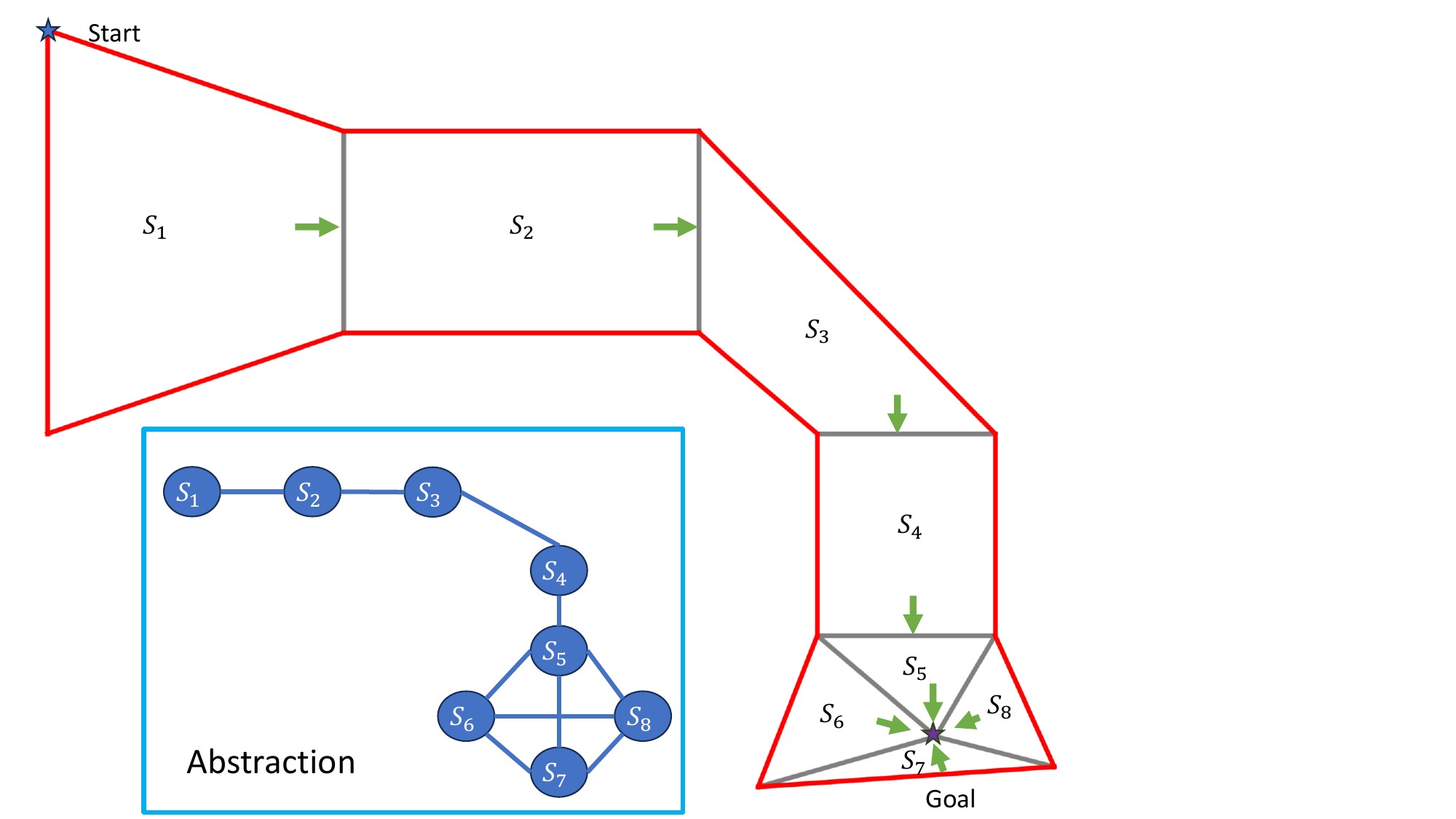}
    \caption{A polygonal environment, shown in red, is partitioned into eight convex regions (cells). Green arrows indicate the exit
direction for each cell. The corresponding high-level planning graph is shown in blue. }
\label{fig:higlevelplanning}
\end{figure}




\subsection{Abstraction and High-level Planning}
This section outlines the process of partitioning a polygonal environment into convex regions and establishing an exit direction for each region.

We consider an agent operating in a polygonal free space $S_W \subset \mathbb{R}^d$. In our setup, we only consider polygonal obstacles. Given $S_W$ with a start point and a goal, we can divide $S_W$ 
into a set of convex polytopes. We denote the $m$-th region by $S_m$.
Each polytope $S_m$ is represented by a set of linear inequalities as:
\begin{equation}
    A_{x,m}x+ b_{x,m} \leq 0 
    \label{eq:edgePolytope}
\end{equation}
Note that these linear inequalities capture the facets of obstacles present in $S_W$ as well. 
These polytopes collectively cover $S_W$ (i.e., $\bigcup_m S_m = S_W$), and their pairwise intersection is at the boundaries, either a plane, segment or vertex.
Without loss of generality, we assume that the goal is located at a vertex of a cell. If this is not the case, we further partition the cell containing the goal (e.g., using triangulation) to get the suitable cells. This partition can be provided by the user or generated using partition algorithms such as in our prior work \cite{mahroorrt}.
To determine the order of cells that must be traversed to reach the goal cell, we abstract the divided environment into a graph $\cG = (\cV, \cE)$ where each node $i\in\cV$ corresponds to cell $S_i$, and each edge $(i,j)\in\cE$ indicates that regions $S_i$ and $S_j$ share a face. We determine a high-level path in $\cG$ from the starting cell to the cell containing the goal using graph search algorithms (e.g.,  Dijkstra \cite{dijkstra1959note}).

Based on the high-level path, we establish the exit direction for each cell. Specifically, the exit direction is orthogonal to the face shared with the subsequent node (i.e., exit face) in the high-level path. In the case of cells containing the goal, the exit face represents a vector pointing directly toward the goal location. See Fig.~\ref{fig:higlevelplanning} for an illustration. 

A similar approach can be used for patrolling tasks; in this case, the high-level path forms a cycle (see \cite{mahroo} for more details).
\label{sec:higlevelplanning}


\section{Problem Formulation}
The goal of this section is to find a controller for each cell that steers the robot toward the exit direction while avoiding other facets of the cell describing the obstacles.

\label{sec:problemformulation}

\subsection{Measurement Model}
The controller needs information regarding the robot's position. We assume that the agent can obtain the PMF of pre-specified landmarks (e.g., using a learning-based sensor). These landmarks can be any distinguishable features in the environment (e.g., the corners of a room). 
Our goal is to construct a stable and safe controller for a large set of PMFs generated by the learning-based sensor. We assume that all landmarks of a cell are always visible to the agent when it is in that cell. We represent $y_i$ as the observation $l_i -x$ of landmark $i$, where $l_i$ is the position of landmark $i$. Its PMF is denoted as $P_{i} \in \mathbb{R}^{n_1 \times n_2 \times \ldots n_d}$, defined over a grid in the reference frame of the robot. We characterize $P_i$ by defining constraints on the mean and Mean Absolute Difference (MAD). We assume that the error between $\mathbb{E}_p[y_i]$ and the ground truth ($l_i -x$) is bounded by $\epsilon \in \mathbb{R}^+$, and the MAD is limited by a user-specified constant $\sigma_m \in \mathbb{R}^+$ in all directions.
Based on these assumptions, the set of all valid realizations of $P_i$ is specified by the following constraints:
\begin{subequations}
 
 \begin{align}
 \label{eq:totalsum0}
 \bold{1}\transpose P_i^{\vee} &= 1\\
 \label{eq:nonneg0}
 P_i &\geq 0 \\
\label{eq:probmean0}
 l_i-x-\epsilon \leq \mathbb{E}_p[y_i-(l_i-x)]&\leq l_i-x+\epsilon \\
\mathbb{E}_p[|(y_i-[l_i-x]_q|]&\leq \sigma_m \quad \forall q \in \mathbb{N}_{\leq d}  \label{eq:probconst0}
\end{align}
\end{subequations}
Note that \eqref{eq:totalsum0} and \eqref{eq:nonneg0} capture the sum of total probability and non-negativity principles.
\begin{remark}
    $\epsilon, \sigma_m$ are lower bounded by resolution of the support $P_i$
\end{remark}
\begin{remark}
We use MAD instead of mean square error (MSE) or variance since the MAD results in an LP formulation. It is possible to use the MSE, but it results in a semi-definite program (SDP) with a large number of variables, which does not scale well as an LP. Using variance leads to a non-convex problem.
\end{remark}

\subsection{ Control Input}
Our goal is to synthesize a linear feedback controller for cell $m$ by using the vectorized version of PMFs ($P_i ^{\vee} \in \mathbb{R}^{n_p}$) of landmarks in cell $m$ as follows:
\begin{equation}
\label{eq:controlcommand}
    u = \sum_{i =1}^{n^m_l} K_{P,i,m} P^\vee_{i,m} +K_{b,m}
\end{equation}

Where $K_{P,i,m} \in \mathbb{R}^{d\times n_p}$ are gain matrices and $K_{b,m} \in \mathbb{R}^d$ is a vector.


\subsection{Problem Statement}
To summarize, given system \eqref{eq:dynamics} and a polytope $S_m$ defined as \eqref{eq:edgePolytope}, derive $K_{P,i,m}$ and $K_{b,m}$ such that the controller \eqref{eq:controlcommand} steers the robot out of the cell in a finite time while avoiding the obstacles captured by \eqref{eq:edgePolytope} and for all possible PMFs specified by \eqref{eq:probconst0}.

\section{Controller Design}
In this part, we first reformulate the probability constraints to better fit our optimization problem. Next, we introduce linear CLFs and CBFs for linear dynamics \eqref{eq:dynamics} and our polygonal environment. We enforce our CLF and CBF constraints by formulating our control problem with a bi-level optimization. Then, by taking the dual of the inner problems and applying robust optimization, we transform the formulation into a regular LP that can be solved efficiently.
In this section, we assume $n^m_l = 1$ for all cells, and we drop index $i$ to simplify the notation. The extension to the multi-landmark scenario will be elaborated upon in Section \ref{sec:multilandmarks}.
\subsection{Reformulation of Probability Constraints}
In this part, we reformulate \eqref{eq:probconst0} constraints to better suit our optimization problem.

Since $\mathbb{E}_p[y_i]$ is a linear operator we can define matrix $U = \bmat{U_1 \\ \vdots\\ U_d}\in \mathbb{R}^{d \times n_p}$ such that $\mathbb{E}_p[y_i] = UP_i^\vee$. Furthermore, we define a vector $z$ to convert absolute constraints (i.e., MAD constraints) to linear constraints as follows:
\begin{equation}
\begin{aligned}
    z_q \transpose P^\vee &\leq \sigma_m\\
    U_q - [l-x]_q \bold{1} & \leq z_q\\
    -U_q + [l-x]_q \bold{1} & \leq z_q\\
    \forall q \in \mathbb{N}_{\leq d}
\end{aligned}
\label{eq:vecVMSE}
\end{equation}

 Note that \eqref{eq:probmean0} is linear in $x$ and $P$; hence, it can be written as the following:
\begin{equation}
\label{eq:probmeanVec}
    A^{\prime}_x x+ A_p P^\vee +b_p \leq 0
\end{equation}


Since we focus on designing the controller for a single cell, we omit index $m$.

\subsection{Linear Control Lyapunov and Barrier Functions}
\label{sec:LCLFCBF}
\subsubsection{Linear Control Lyapunov Function}

 We desire to define a CLF ($V$) for each cell to ensure that the agent will reach the exit face in a finite time regardless of the starting point. This can be achieved by the following linear CLF :
\begin{equation}
\label{eq:lclf}
    V(x) = v \transpose (x - o) 
\end{equation}
Where $v$ is the negation of the exit direction (i.e., inward-facing normal to the exit face)  $o$ is a point on the exit face. Note that $V(x)$ is positive for any point in the region and zero for any point on the exit face. Thus, this $V(x)$ is a valid CLF for our goal.\\
Applying \eqref{eq:clfdef} for the CLF constraint leads to
\begin{equation}
\label{eq:lclfdot}
\tag{CLF}
   v\transpose  (Ax+B(K_P P^\vee+K_b) ) +\alpha_v v\transpose (x-o)  \leq 0
\end{equation}
\subsubsection{Linear Control Barrier Function (CBF)}
We specify our safe set by $n_h$ linear constraints, that capture all of the obstacle facets of $S$. Hence, we can define $n_h $ CBFs as follows:
\begin{equation}
\label{eq:lcbf}
    [h]_j = [A_{h}]_j x +[b_{h}]_j \quad \forall j \in {N}_{\leq n_h}
\end{equation}
\begin{remark}
    $V(x)$ and $[h]_j$ are equal to the normal distance to the exit face and wall $j$, respectively.
\end{remark}
By substituting \eqref{eq:cbfdef} CBF constraint \eqref{eq:lcbf} we obtain:
\begin{equation}
\label{eq:lcbfdot}
\tag{CBF${}_j$}
\begin{aligned}
      &- [A_{h}]_j(Ax+B(K_{p}P^\vee+K_b)) -\alpha_h( [A_{h}]_j x +[b_{h}]_j)\leq 0  \\&\quad \quad \quad \quad \forall j \in {N}_{\leq n_h}
\end{aligned}
\end{equation}
\subsubsection{Matrix Form}
Since CLF and CBF constraints are affine in $x$ and $P$, we introduce the following concise format for \eqref{eq:lclfdot} and \eqref{eq:lcbfdot} :
\begin{equation}
\begin{aligned}
    &{c_{x}^k} \transpose x_+ {c_{p}^k }\transpose P^{\vee}+r^k \leq 0 \quad \forall k \in \mathbb{N}_{\leq n_h +1}
\end{aligned}
\end{equation}
Where $k=1$ corresponds to \eqref{eq:lclfdot} and $1 < k \leq n_h +1$ corresponds to \eqref{eq:lcbfdot}.
We seek to satisfy such constraints for all $x\in S$  and $ z, P \in \cF (x)$, where $\cF (x)$ denotes the feasible set for $z$ and $P$ for a given $x$. $\cF (x)$ is characterized by \eqref{eq:totalsum0}, \eqref{eq:nonneg0} ,and \eqref{eq:vecVMSE} constraints.

\begin{remark}
This approach can be extended to higher-order systems using higher-order CLF and CBF \cite{xiao2021high}. But in this case, the feasible set of $x$ would depend on the initial condition.
\end{remark}

\subsection{Control Synthesize Problem}
In this section, we formulate an optimization problem formulation to synthesize the controller such that enforces \eqref{eq:lclfdot} and \eqref{eq:lcbfdot} for all points of the cells and all realizations of $P$. 
We formulate the following optimization problem :
\begin{equation}
\label{eq:primalopt}
    \begin{aligned}
        \max_{\delta_k, K, K_{b} } \quad & \sum_{ k =1}^{n_h+1} \omega_k \delta_k \\
          &\subjectto  
          \left[ \begin{aligned}
              &{c_{x}^k} \transpose x + {c_{p}^k} \transpose P^\vee +r^k\\
               &\forall x \in S , P,z \in \cF (x)
          \end{aligned} 
          \right] \leq -\delta_k\\
        & \delta_k \geq 0, \quad \forall  k \in \mathbb{N}_{\leq n_h +1}
    \end{aligned}
\end{equation}
Intuitively, the constraints indicate that the CLF and CBF constraints must be satisfied with margins $\delta_k$. The objective is to maximize these margins, thereby achieving the most robust control gains. Eq \eqref{eq:primalopt} has an infinite number of constraints. We address this in two steps. First, for a given $x$ and $z$, we seek to satisfy the maximum of each constraint over $P$. Therefore, \eqref{eq:primalopt} becomes a bi-level optimization as follows :
\begin{equation}
\label{eq:primal_innervvv}
\begin{aligned}
     &\max_{\delta_k, K, K_{b} } \quad  \sum_{ k =1}^{n_h+1} \omega_k \delta_k \\
     &\subjectto\\
       &\left (\begin{aligned}
    &\max_{P} \quad   { c_p^k} \transpose P ^ \vee\\
        &\subjectto \\
        & \bold{1} \transpose P^\vee -1 = 0 & : \lambda_s^k \\
        & A_p P^\vee + A^\prime_x x+ b_p  \leq 0  & : \lambda_{p}^k \\
        &z_q \transpose P ^\vee \leq \sigma_m &: \lambda_{zq}^k \\
        &P \geq 0
    \end{aligned} \  
    \right)\begin{aligned}
    \\
     \leq \begin{aligned}
     \\
         &-\delta_k-r^k-{c^k_x }\transpose x\\ 
         &\forall x \in S, z \in \cF(x)
     \end{aligned}
    \end{aligned}\\
     & \delta_k \geq 0, \forall k \in \mathbb{N}_{\leq n_h+1}, q \in \mathbb{N}_{\leq d} 
\end{aligned}
\end{equation}
We take the dual of the inner problems to convert it to a minimization problem where $\lambda_s^k, \lambda_p^k, \lambda_z^k$ are the corresponding dual variables denoted in \eqref{eq:primal_innervvv}. The dual of the inner problem is equal to:
\begin{equation}
\label{eq:innerdual}
    \begin{aligned}
        \min_{\lambda_s^k, \lambda_p^k, \lambda_{zq}^k} & - {\lambda_p^k} \transpose A^\prime_x x - {\lambda_p^k} \transpose b_p + \lambda_s^k + \sum_{q=1}^d \lambda_{zq}^k \sigma_m \\
        & \subjectto 
        -{c_p^k} \transpose+ {\lambda_p^k} \transpose A_p + \lambda_s^k \bold{1} \transpose +\sum_{q=1}^d \lambda_{zq}^k z_q\transpose \geq 0 \\
        &\lambda_p^k, \lambda_{zq}^k \geq 0
    \end{aligned}
\end{equation}
Due to strong duality (i.e., the optimal objective value of the dual and primal are equal.) of LPs, we replace the inner problems with their duals. This allows us to convert the problem to :

\begin{equation}
\label{eq:ControlDesignD1}
    \begin{aligned}
       & \max_{K,K_b, \delta_{k}, \lambda_{s,x,z}^k } \quad \sum_{k =1}^{n_h+1} \omega_k \delta_k\\
        &\subjectto\\
          & - {\lambda_p^k} \transpose A^\prime_x x - {\lambda_p^k} \transpose b_p + \lambda_s^k + \sum_{q=1}^d \lambda_{zq}^k \sigma_m  \leq -\delta_k-r^k-{c^k_x} \transpose x\\
        &-{c^k_p} \transpose+ {\lambda_p^k} \transpose A_p + \lambda_s^k \bold{1} \transpose +\sum_{q=1}^d \lambda_{zq}^k z_q\transpose \geq 0 \\
      &\lambda_p^k, \lambda_{zq}^k, \delta_k \geq 0 \quad
        \forall x \in \cX , p,z \in \cF (x), \forall k \in \mathbb{N}_{\leq n_h +1}
    \end{aligned}
\end{equation}

Eq \eqref{eq:ControlDesignD1}  still has infinite constraints, due to considering all possible $x \in S$ and $z \in \cF(x)$. To address this, we repeat the same procedure as before, satisfying the maximum of constraints over $x \in S$ and $z \in \cF (x)$. This leads to the following bi-level optimization problem:
\begin{equation}
    \begin{aligned}
    &\max_{K,K_b, \delta_{k}, \lambda_{s,x,z}^k } \quad\sum_{k =1}^{n_h+1} \omega_k \delta_k\\
    &\subjectto\\
         &\left (\begin{aligned}
   &\max_{x, z} {c_x^k}\transpose x - {\lambda_p^k} \transpose A_x ^\prime x\\
        &\subjectto \\
        & A_x x + b_x \leq 0 & : \lambda_x^k \\
        & U_q- e_q\transpose(l-x) \bold{1} \leq z_q & : \rho_{q1}^k \\
        &-U_q+ e_q\transpose(l-x) \bold{1}\leq z_q & : \rho_{q2}^k \\
        & \forall q \in \mathbb{N}_{\leq 2}
    \end{aligned} \  
    \right) \leq \begin{aligned} \\ \\
       & {\lambda_p^k} \transpose b_p -r^k -\lambda_s^k \\& - \sum_{q=1}^d \sigma_m \lambda_{zq}^k-\delta_k 
    \end{aligned}\\
     &\left (\begin{aligned}
   &\max_{x, z} -\sum_{q=1}^d \lambda_{zq}^k [z_q]_i \\
        &\subjectto \\
        & A_x x + b_x \leq 0 & : [\beta_x]_i^k \\
        & [U_q]_i- [l-x]_q \leq [z_q]_i & : [\eta_{q1}]_i^k \\
        &-[U_q]_i+ [l-x]_q \leq [z_q]_i & : [\eta_{q2}]_i^k \\
        & \forall q \in \mathbb{N}_{\leq d}
    \end{aligned} \  
    \right)\leq \begin{aligned} \\
        &[-{c_p^k} \transpose+ {\lambda_p^k} \transpose  A_p\\& + \lambda_s ^k \bold{1} \transpose]_i
    \end{aligned} \\
        & \lambda_p^k , \lambda_z^k,\delta_k \geq 0 \quad \forall k \in \mathbb{N}_{\leq n_h +1} \, \forall i \in \mathbb{N}_{\leq n_p}
    \end{aligned}
\end{equation}
Similarly, we take the dual of the inner problem. The dual of the first problem equals to:
\begin{equation}
    \begin{aligned}
        &\min_{\lambda_x^k, \rho^k}  -\lambda_x^kb_x+ \sum_{q=1}^d (\rho_{q1}^k-\rho_{q2}^k)(-U_q +[l]_q \bold{1}) \\
        &\subjectto
        \rho_{q1}^k +\rho_{q2}^k =0 \\
         &-{c_x^k}\transpose + {\lambda_p^k} \transpose A^{\prime}_x + \sum_{q=1}^d {(\rho_{q1}^k-\rho_{q2}^k)}\transpose \bold{1} e_q\transpose+ {\lambda_x^k} \transpose A_x = 0 \\
        & \lambda_x^k, \rho_{q1}^k, \rho_{q2}^k \geq 0  
    \end{aligned}
\end{equation}
Taking the dual of second problem results in:
\begin{equation}
    \begin{aligned}
        \min_{\beta_x^k, \eta^k_i} & - {\beta_x^k} \transpose b_x+ \sum_{q=1}^d  ([\eta_{q1}]^k_i-[\eta_{q2}]^k_i)(-[U_{q}]_i+l_q)  \\
        & \subjectto \lambda_{zq}^k - [\eta_{q1}]^k_i -[\eta_{q2}]^k_i = 0 \quad 
        \forall q \in \mathbb{N}_{\leq d}\\
        & {\beta_x^k} \transpose A_x+ \sum_{q=1}^d[\eta_{q1}^k-\eta_{q2}^k]_i e_q \transpose = 0
    \end{aligned}
\end{equation}

Applying robust optimization simplifies our control synthesis problem into a regular LP
as shown below:
\begin{equation}
\label{eq:finalopt}
\begin{aligned}
   & \max_{K, K_b, \delta_h, \delta_v, \lambda_{p,s,x}^{k}, \rho^k, \beta_x^k }   \sum_{k =1}^{n_h+1} \omega_k \delta_k \\
    & \text{subject to} \\
    &\forall k \in \mathbb{N}_{\leq n_h+1} \quad \left\{
    \begin{aligned}
    & - {\lambda_{x}^k} \transpose b_x + \sum_{q=1}^d{(\rho^k_{q1}-\rho^k_{q2})}\transpose (-U_q +[l]_q \bold{1})\\
    & - {\lambda_p^k} \transpose b_p + {\lambda_z^k}\transpose \bold{1} \sigma_m + \lambda_s^k \leq -\delta_k - r^k \\
    & - {c_{x}^k} \transpose + {\lambda_{p}^k} \transpose A_x^\prime + \sum_{q=1}^d (\rho_{q,1}^k-\rho_{q,2}^k) \bold{1} e_q \transpose = 0 \\
    & \rho_{q1}^k+\rho_{q2}^k = 0 \quad \forall q \in \mathbb{N}_{\leq d} \\
    & - \beta_x^k b_x + \sum_{q=1}^d (\eta_{q1}^k-\eta_{q2}^k)\odot(-U_q + l_q \bold{1})\\
    & \leq -c_p^k  + A_p\transpose  \lambda_p^k + \lambda_s^k \bold{1}\\
    & \beta_x^k A_x + \sum_{q=1}^d(\eta_{q1}^k-\eta_{q2}^k)e_q\transpose = 0\\
    & \lambda_{zq}^k \bold{1} - \eta_{q1}^k - \eta_{q2}^k = 0 \quad \forall q \in \mathbb{N}_{\leq d}\\
    & \lambda^{k}_{z,x,p}, \rho^{k}_q, \eta^{k}_q, \beta_x^{k}, \delta_k \geq 0  
    \end{aligned} \right.
\end{aligned}
\end{equation}

\begin{remark}
In the case of the cell containing the goal, when the agent reaches its destination, $u$ must be reduced to zero. However, this condition is not inherently encoded \eqref{eq:controlcommand}. To address this, we can add $u_{goal} = K_{P} P_{goal}^\vee +K_{b} = 0$ to \eqref{eq:finalopt} as a constraint. $P_{goal}$ is the delta PMF  where the only point corresponding to the grand truth is one and others are zero of the landmark when the agent is at the goal. 
\end{remark}

\begin{figure*}
  \centering

  \subfloat[$\epsilon =2\,\sigma_m = 9$, delta ]{%
    \includegraphics[width=0.24\textwidth]{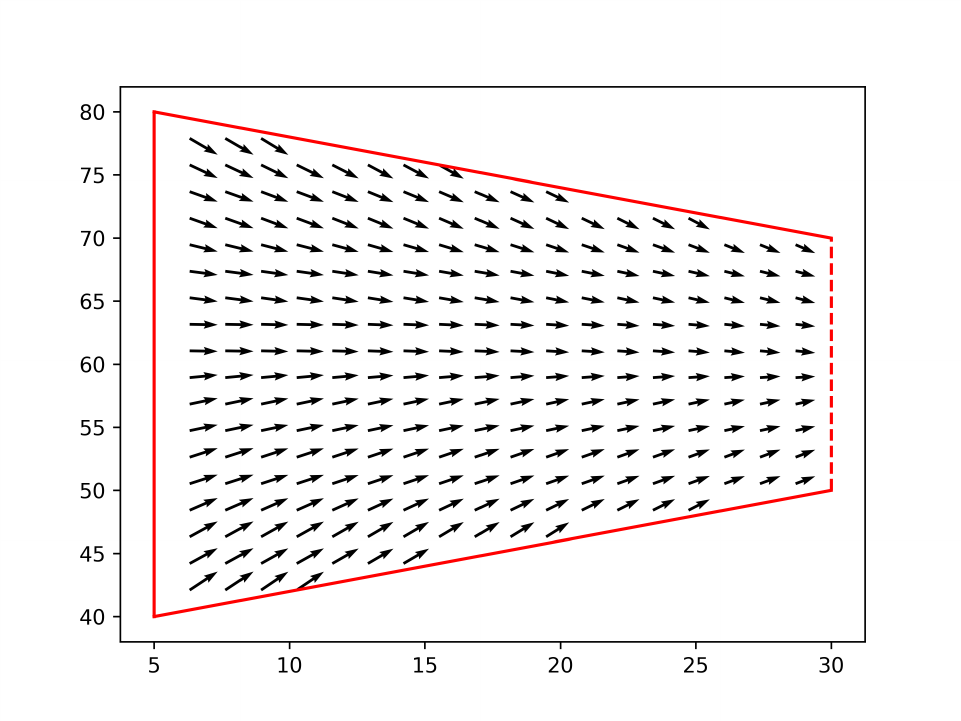} \label{fig:env}
  }
  \hfill
  \subfloat[$\epsilon =8\,\sigma_m = 128$, delta ]{%
    \includegraphics[width=0.24\textwidth]{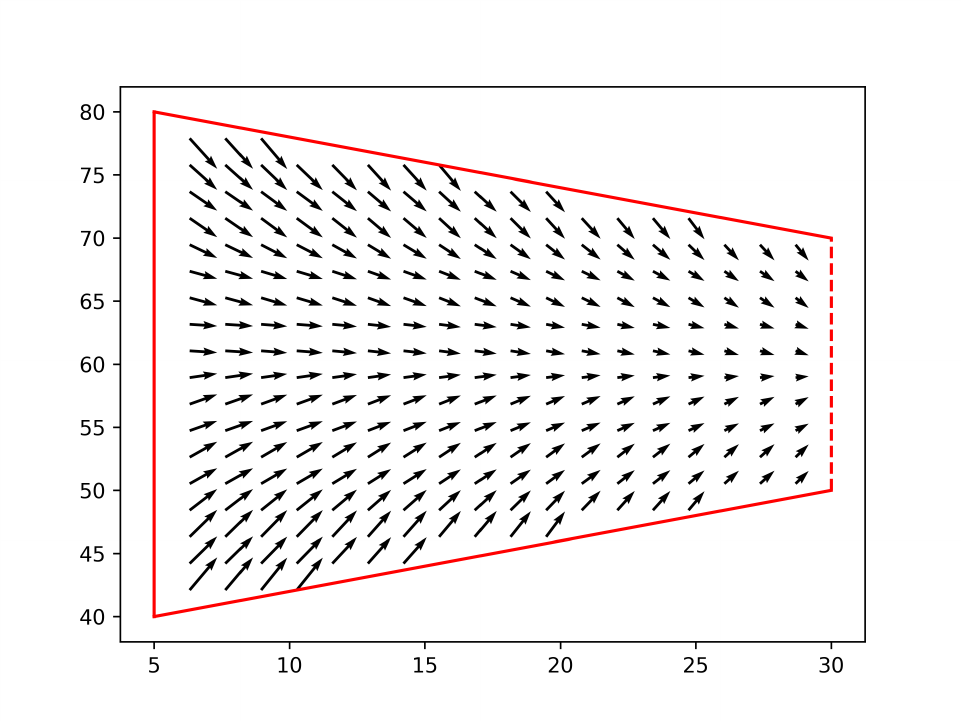}\label{fig:envdc}
    }
  \hfill
  \subfloat[$\epsilon =2\,\sigma_m = 9$, Gaussian ]{%
    \includegraphics[width=0.24\textwidth]{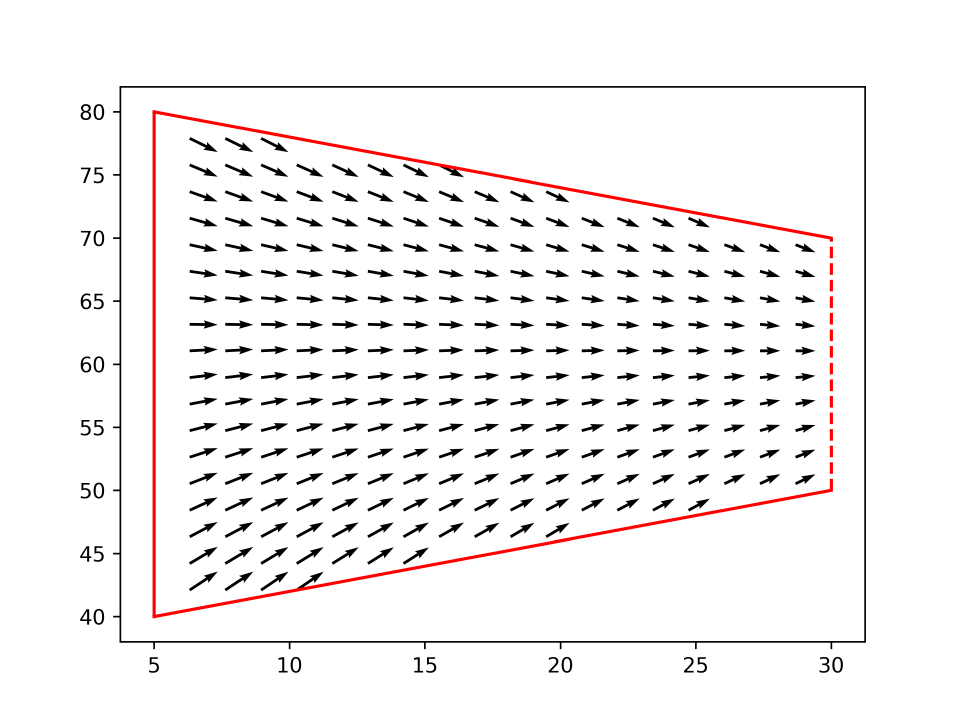}\label{fig:envabstract}
  }
   \hfill
  \subfloat[$\epsilon =8\,\sigma_m = 128$, Gaussian ]{%
    \includegraphics[width=0.24\textwidth]{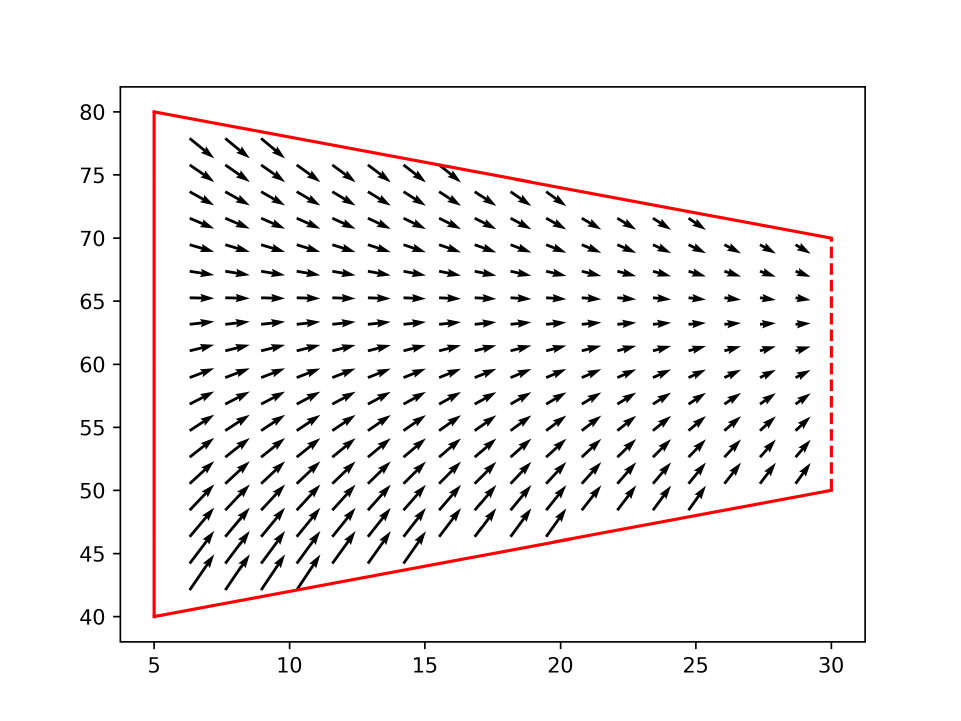}\label{fig:envabstract}
  }

  \caption{Vector field of a single cell for different hyperparameters and measurement models.  Dashed segments show the exit facets. More uncertainty ( larger $\sigma_m, \epsilon$ ) leads to a more conservative and robust controller. }
  \label{fig:vecf}
\end{figure*}
\subsubsection{Regularization of the Controller}
  $K_p \in \mathbb{R}^{d\times n_p}$ is a matrix where each row is of the same size  $P$. In order to reduce the number of decision variables in \eqref{eq:finalopt}, we can parameterize $K_p$ with a set of constant gain $K_i \in \mathbb{R}^{d \times d}$ and a set of  linear maps $R_i \in \mathbb{R}^{d \times n_p}$ :
\begin{equation}
 K_p = \sum_{i=1}^{n_k} K_i R_i
\end{equation}
 $R_i$ represents a design choice affecting the controller; a convenient choice is to use a discretized version of a function over the support of $P$, (e.g.,  quadratic, sinusoid, etc.). Moreover, this allows employing the same controller for a smaller grid size.
\paragraph{Relation with \cite{mahroo}}
In our previous work \cite{mahroo}, we used $u=K Y$,
where $Y$ is the estimated positions of landmarks in the PMF $P$. One way to obtain $Y$ is to take the empirical mean resulting in $u=K U P^\vee$.

\subsection{Multiple landmarks}
\label{sec:multilandmarks}
This approach can be extended to multiple landmarks. In this case, each PMF of landmark $i$ is characterized by \eqref{eq:probconst0}. We stack all 
PMFs into one matrix denoted as
$P_\text{all}= \bmat{ P_1^\vee &\ldots P_{n_l^m}^\vee } $.
Note that CBF and CLF remain affine with respect to $P_\text{all}$, thus we can use our matrix format and represent CLF and CBF constraints as follows:
\begin{equation}
     {c_{x}^k} \transpose x_+ {c_{p_{\text{all}} }^k} \transpose P^\vee_{\text{all}}+r^k \leq 0 \quad \forall k \in \mathbb{N}_{\leq n_h +1}
\end{equation}
Therefore, we can adapt our bi-level optimization \eqref{eq:primal_innervvv} by replacing $P$ with $P_\text{all}$. By applying the same approach(using robust optimization and taking duals), the modified problem can be converted to an LP.

\section{Case Study}

\begin{figure}
\centering

     \includegraphics[width=0.48\textwidth]{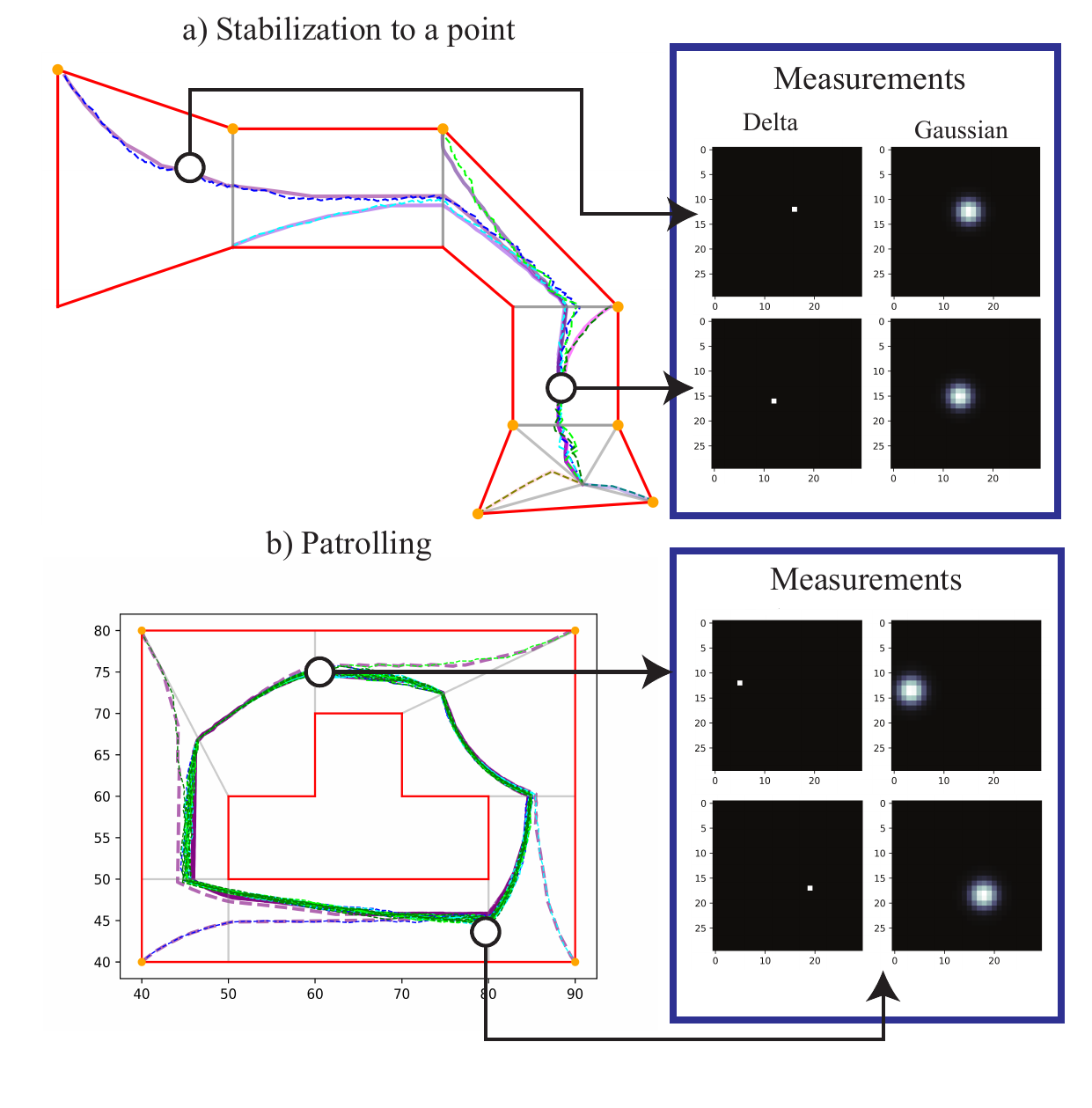}

  \caption{ Simulating for two tasks: patrolling and stabilizing. Red lines demonstrate walls and gray segments represent exit faces. Yellow circles indicate landmarks. We generated various trajectories, starting from different points, considering both delta and Gaussian PMFs represented by solid and dashed curves, respectively. Two examples of delta and Gaussian measurements for two distinct points used by the controller are shown.  Our approach effectively handles relatively large drift and variance errors.   }
  \label{fig:traj}
\end{figure}

We implement the single-landmark algorithm for a first-order integrator system for stabilizing and patrolling tasks in 2D using the following hyperparameters: $\alpha_h = 100$, $\alpha_v = 1$, $\epsilon = 4, n_1,n_2 = 30$, and $\sigma_m = 16$. We used the corners of environments as landmarks. We parameterize our control gains using mean, quadratic, and cosine linear maps. We assess our controller for both the delta PMFs (where only the point corresponding to ground truth equals one and others are zero) and Gaussian PMFs. The Gaussian PMFs are generated by convoluting the delta PMFs with a kernel having a drift error of $3$ and a variance of $\sigma^2 =12$. 
 The resulting trajectories for both cases are shown in Figure \ref{fig:traj}. Notably, the controller successfully completes tasks in scenarios with both certain and uncertain measurements.
To further evaluate the robustness of our approach, we plot the vector field generated by our controller within a single cell for varying values of $\sigma_m$ and $\epsilon$ shown in Figure \ref{fig:vecf}. As demonstrated in our case studies, the controller exhibits significant robustness, which is not typically expected from a fully linear system. 

\section{Conclusion}
Overall, our proposed method offers a novel solution for generating linear feedback controllers that can operate within polygonal environments, providing robustness to mapping and measurement variations while ensuring safety and stability. The set of control gains for the entire environment is determined based on LP using CLF and CBF constraints, which are solved offline. The controller uses the entire PMFs, which leads to a robust controller with minimal filtering and preprocessing. We validated the performance of this approach through a simulation for two case studies. Because of minimal online computation, this controller can be applied to low-budget systems.  \\
For future work, we seek to extend the method to higher-order systems and consider a limited field of view for landmarks. 
Additionally, we plan to implement the entire pipeline, including the learning-based perception module, on real hardware.
\bibliographystyle{biblio/ieee} 
\bibliography{main} %

\end{document}